\documentclass[preprint,
 amsmath,amssymb,
 aip,
jap
]{revtex4-1}

\usepackage{CJK}

\usepackage{graphics}
\usepackage{graphicx}
\usepackage{amssymb}
\usepackage{amsmath}
\usepackage{subfigure}
\usepackage{siunitx}

\providecommand*{\sub}[1]{\ensuremath{_\mathrm{#1}}}

\newcommand{\kp}{\mbox{\ensuremath{\boldsymbol{k}\!\cdot\!\boldsymbol{p}}}}

\newcommand{\Vg}{\ensuremath{V \sub G}} 
\newcommand{\Vt}{\ensuremath{V \sub T}}

\newcommand{\zi}{\ensuremath{x\sub i}}

\newcommand{\Cinv}{\ensuremath{C\sub{inv}}}
\newcommand{\Cc}{\ensuremath{C\sub{c}}}
\newcommand{\Cq}{\ensuremath{C\sub{q}}}
\newcommand{\Cins}{\ensuremath{C\sub{ins}}}

\newcommand{\phis}{\ensuremath{\psi\sub{s}}}
\newcommand{\phic}{\ensuremath{\psi\sub{c}}}

\newcommand{\Tins}{\ensuremath{T\sub{ins}}}

\newcommand{\epsilons}{\ensuremath{\varepsilon\sub{s}}}

\newcommand{\Qi}{\ensuremath{Q\sub{i}}}

\newcommand{\Cg}{\ensuremath{C\sub G}}

\newcommand{\centroid}{\ensuremath{\Delta}}

\newcommand{\DeltaSO}{\Delta\sub{SO}}
\newcommand{\dd}{\,\text{d}}
\newcommand{\eqdot}{\,\text{.}}

\begin{document}
\pdfoutput=1 
\title{Electrostatic Performance of InSb, GaSb, Si and Ge p-channel Nanowires}

\author{Celso Martinez-Blanque}
\affiliation{Dpto. de Electr\'onica y Tecnolog\'ia de Computadores, Facultad de Ciencias, Universidad 
de Granada. Av.Fuentenueva S/N, 18071 Granada, Spain}
\author{Enrique G. Marin}
\affiliation{Dipartimento di Ingegneria dell\'{}Informazione, Universit\`{a} di Pisa, Via G. Caruso 16, 56122 Pisa, Italy}
\author{Alejandro Toral}
\author{Jos\'e M. Gonzalez-Medina}
\author{Francisco G. Ruiz}
\email{franruiz@ugr.es}
\author{Andr\'es Godoy}
\email{agodoy@ugr.es}
\author{Francisco G\'{a}miz}
\affiliation{Dpto. de Electr\'onica y Tecnolog\'ia de Computadores, Facultad de Ciencias, Universidad 
de Granada. Av.Fuentenueva S/N, 18071 Granada, Spain}

\begin{abstract}

The electrostatic performance of p-type nanowires (NWs) made of InSb and GaSb, with special focus on their gate capacitance behavior, is analyzed and compared to that achieved by traditional semiconductors usually employed for p-MOS such as Si and Ge. To do so, a self-consistent $\kp$ simulator has been implemented to achieve an accurate description of the Valence Band and evaluate the charge behavior as a function of the applied gate bias. The contribution and role of the constituent capacitances, namely the insulator, centroid and quantum ones are assessed. It is demonstrated that the centroid and quantum capacitances are strongly dependent on the semiconductor material. We find a good inherent electrostatic performance of GaSb and InSb NWs, comparable to their Ge and Si counterparts making these III-Sb compounds good candidates for future technological nodes. 



\end{abstract}
\maketitle


\section{Introduction}\label{sec:introduction}

In the downscaling roadmap of Metal-Oxide-Semiconductor Field-Effect-Transistors (MOSFETs), industry has already reached the milestone of transition from planar to three-dimensional devices enforced by the necessity of controlling the Short Channel Effects (SCEs). In this context, nanowires (NWs) have been postulated as the most efficient structure to reduce SCEs as they provide a higher gate electrostatic control of the channel compared to other multigate devices. In addition, the requirements of both high performance and low power devices demand the use of materials different to Si, able to reduce the supply voltage while maintaining or increasing the on current ($I\sub{ON}$). 

In this scenario, and due to their high bulk mobility, \mbox{III-V} materials have attracted extensive research interest in recent years \cite{DelAlamo2011}, being recognized as promising building blocks for the next generation of electronics and photonics \cite{Riel2014}. Most of current works in the literature focus on n-type devices, where materials such as GaAs, InAs and InGaAs have already demonstrated impressive performance. Unfortunately, the enhancement in the electron mobility of these materials is not accompanied by a similar enhancement in the hole mobility, resulting in a poorer performance for pMOS transistors. 
To achieve fully operational complementary-MOS (CMOS) circuits, the tolerable asymmetry between both transistors cannot be very marked, and thus, the search for a high performance nMOS is bound to find its equivalent pMOS. Several materials are currently being investigated as technologically feasible p-type channels, being Ge one of the preferred choices due to its high bulk mobility \cite{Pillarisetty2011}. Nonetheless, increasingly more attention has been focused on antimonide compounds, such as InSb and GaSb, owing to their excellent bulk hole mobilities among the rest of \mbox{III-V} compounds \cite{Wernersson2013}. Particularly, GaSb has demonstrated the ability to provide both, high bulk hole and electron mobility, $\sim$\SI{1400}{\square\centi\meter\per\volt\per\second} and \SI{7000}{\square\centi\meter\per\volt\per\second}, respectively \cite{Shimoida2013}.

\begin{figure}
\centering
\includegraphics[width=50mm]{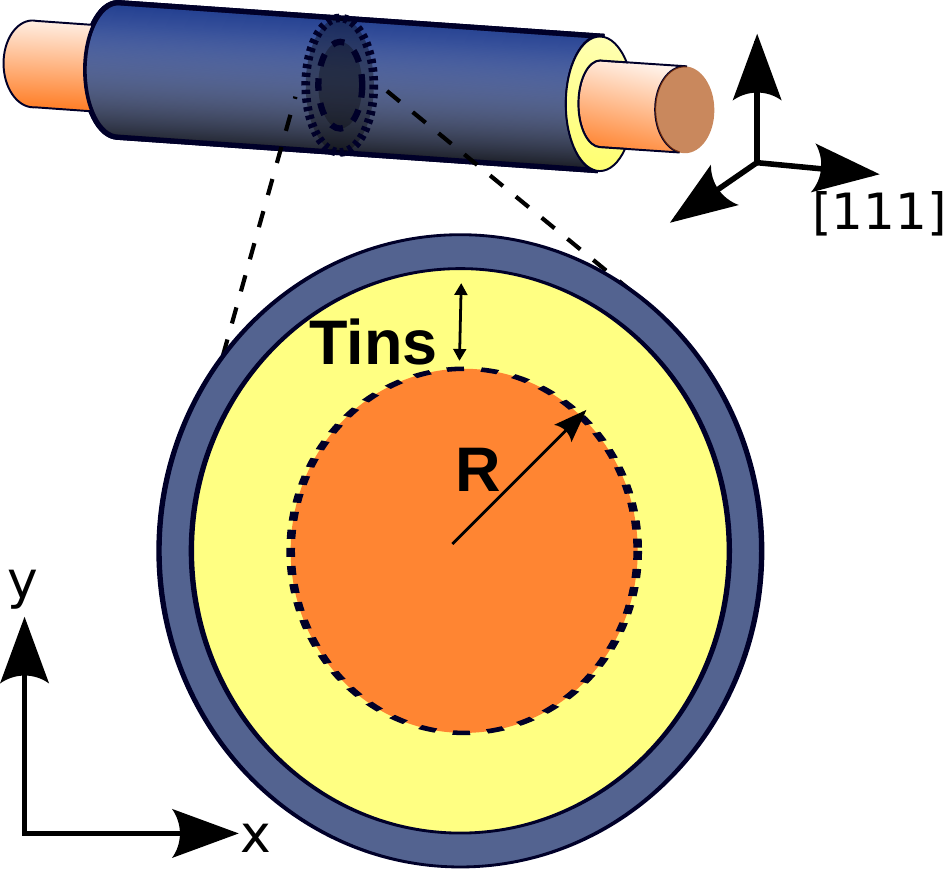}
\caption{Geometry of the cylindrical NW, where T$_{\mathrm{ins}}$ is the gate insulator thickness and R the semiconductor radius.}\label{fig:nanohilo}
\end{figure}

A high mobility, however, does not guarantee a good performance. A potential substitute of silicon must also guarantee a large charge modulation in the channel, namely a good electrostatic control of it. This characteristic of a device is mainly determined by the gate capacitance and the inversion charge. In this sense, \mbox{III-V} materials have been largely accused to be affected by a Density of States (DOS) bottleneck. 
However, there are few studies dealing with this issue and its impact on the overall performance of pMOS NWs.

To this purpose, in this work we address the electrostatic properties and gate capacitance of GaSb and InSb NWs, and compare them to that achieved by Si and Ge devices, assessing whether the expected high mobility of these compound semiconductors may be degraded by a poorer electrostatic performance.\\


The outline of the paper is as follows. First, Section \ref{sec:numerical simulator} describes the numerical simulator developed for this study and the main parameters employed to model each material. Next, in  Section \ref{sec:results and discussion}, the main results regarding the bandstructure, charge distribution and gate capacitance for the different NWs are shown and analyzed, focusing on the main differences between the four considered materials. That Section also provides insights on the influence of the insulator capacitance, which is modified by changing the dielectric material. Finally, the main conclusions of this work are drawn in Sec. \ref{sec:conclusion}.


\section{Numerical Simulator} \label{sec:numerical simulator}

\begin{table}[b]
\centering
\begin{tabular}{l|cccc}
\hline
  &\textbf{Si} & \textbf{Ge} & \textbf{GaSb} & \textbf{InSb}\\
 \hline
 \hline $E\sub g\;$(\si{\electronvolt})  & $1.17$   & $0.742$ & $0.812$   & $0.235$\\
 $\DeltaSO\;$(\si{\electronvolt}) & $ 0.044$ & $0.286$ & $0.76$    & $0.81$\\
 $E\sub P \;$(\si{\electronvolt})  & $-$      & $-$     & $18.0$    & $23.3$\\
 $m\sub c \;$  & $-$      & $-$     & $0.039$   & $0.0135$\\
 $\gamma_1$ & $3.55$   & $13.27$ & $13.4$    & $34.8$ \\
 $\gamma_2$ & $0.65$   & $4.32$  & $4.7$     & $15.5$\\
 $\gamma_3$ & $1.26$   & $5.61$  & $6.0$     & $16.5$ \\
 $\varepsilon\sub r$ & $11.7$ & $16.2$ & $15.7$ & $16.8$ \\
 $\Xi_0 \;$ (\si{\electronvolt}) & $ 4.05$     & $4.00$  & $4.06$    & $4.59$
\end{tabular}
\caption{$\kp$ parameters for Si \cite{Shin2009}, Ge \cite{Fischetti1996}, GaSb and InSb \cite{Vurgaftman2001}. 
}
\label{tab:kp_parameters}
\end{table}

In this study we are interested in the electrostatic behavior of cylindrical gate-all-around (GAA) p-type NWs made of four different materials: two \mbox{III-V} antimonides, GaSb and InSb, and two group IV semiconductors, Si and Ge. Particularly, we have studied NWs with 5nm diameter oriented along the [111] direction (Fig. \ref{fig:nanohilo}), since this orientation exhibits the best transport properties for holes in Si and Ge \cite{Neophytou2013c,Niquet2012d,Martinez-Blanque2014}. To allow a fair comparison of the electrostatics, the same high-$\kappa$ insulator, $\mathrm{Al}_2\mathrm{O}_3$, with thickness $\Tins=1.5$nm and relative dielectric constant $\varepsilon\sub r=9$, is considered in all cases. This way, the contribution of the insulator capacitance (\Cins) is identical for all the NWs and the comparison can be focused on the channel material.

In order to describe the electrical performance of the semiconductor NWs under study, we have used a self-consistent solver for the two-dimensional (2D) Schr\"odinger and Poisson equations in the transversal cross-section of the cylindrical GAA NWs. The effective mass approximation with non-parabolic corrections has demonstrated to provide a simple but accurate description of the band structure of \mbox{III-V} compounds close to the Conduction Band (CB) minimum and has, therefore, been used in similar studies of n-type NWs \cite{Marin2015}. However, this method fails in the description of the more complex Valence Band (VB) of group IV (Si, Ge) and \mbox{III-V} semiconductors due to the coupling of the different subbands encompassing the VB: Heavy Holes (HH), Light Holes (LH) and Split-Off (SO). Thus, to achieve an accurate description of the VB of such materials, it is mandatory to use a model which accounts for the coupling between these subbands. 

In this work, the $\kp$ method, which explicitly accounts for the coupling between VB subbands has been implemented to calculate the VB structure and the associated envelope wave functions \cite{Lassen2006,Shin2010,RaviKishore2012}. The $\kp$ simulation scheme employs a reduced set of parameters which can be obtained semi-empirically as they are tuned with other atomistic descriptions. The results presented in this work have been benchmarked against atomistic methods, such as Tight-Binding (TB), providing a good agreement around the band edges \cite{Shin2009,Shin2010}. For indirect gap materials, such as Si and Ge, a six-band model is accurate enough since the different subbands forming the VB are considered explicitly, whereas the rest of bands, including the CB, are regarded as remote. However, for direct materials with a small gap, such as the \mbox{III-Sb} compounds here analyzed, the proximity of the CB to the VB has a non negligible influence on the resulting VB structure. The CB can also be explicitly accounted for in an eight-band $\kp$ model \cite{Bahder1990}. The implemented model uses a set of semi-empirical parameters directly related to the different anisotropic effective masses of the involved subbands, the so-called Luttinger parameters, that can be found in the literature \cite{Vurgaftman2001}. These parameters are valid for both, six-band and eight-band $\kp$ models. However, for the latter, they have to be reduced to exclude the effect of the CB, which is explicitly taken into account in the model \cite{Bahder1990}. The $\kp$ parameters used in this study, and the references from where they have been extracted, are presented in Table \ref{tab:kp_parameters}. The Kane coupling parameter $E\sub P$ has been reduced for GaSb, as suggested in \cite{Veprek2009}, to avoid the appearance of spurious solutions.



\section{Results and Discussion}\label{sec:results and discussion}

\subsection{Valence Band bandstructure}

\begin{figure}
\centering
\includegraphics[width=80mm]{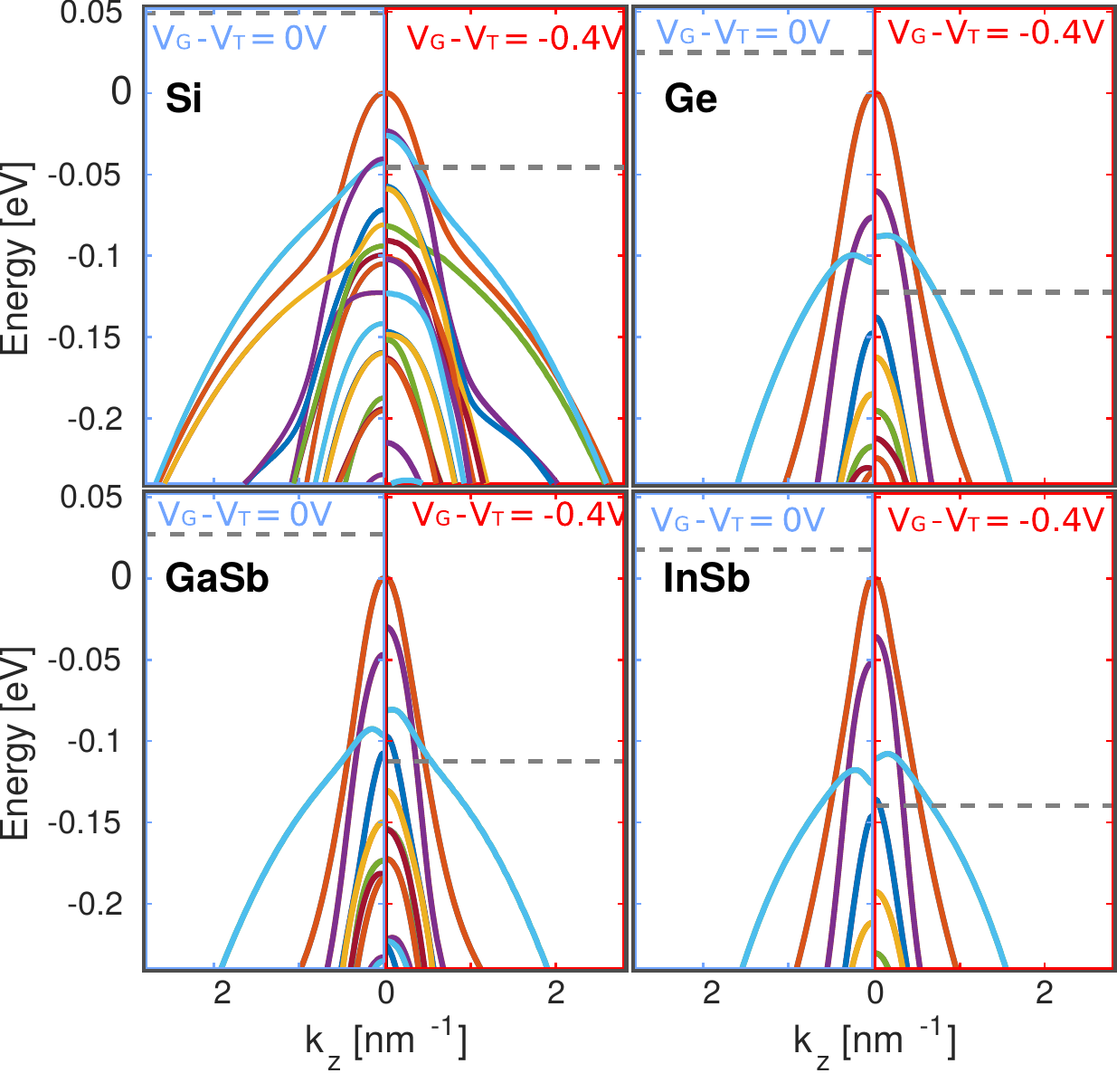}
\caption{Bandstructure for 5nm diameter cylindrical NWs with channel materials: Si (upper left), Ge (upper right), GaSb (bottom left) and InSb (bottom right), at a gate overdrive voltage of 0V (left) and \SI{-0.4}{\volt} (right). The energy is referred to the VB edge, and the Fermi level is depicted as a dashed line.}\label{fig:bandstructure}
\end{figure}

\begin{figure}
\centering
\includegraphics[width=80mm]{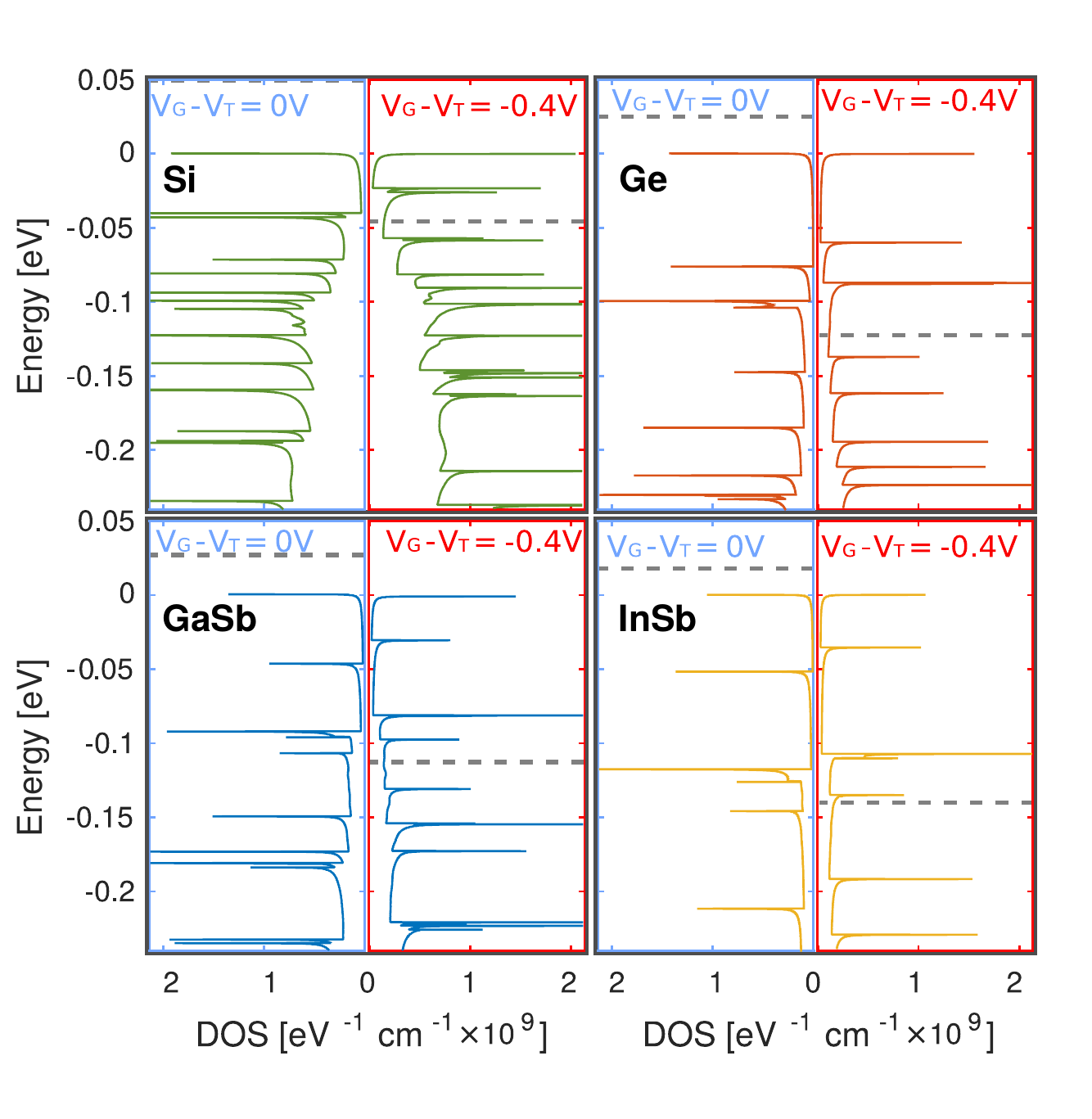}
\caption{DOS evaluated for 5nm diameter cylindrical NWs with four different channel materials: Si (upper left), Ge (upper right), GaSb (bottom left) and InSb (bottom right), at a gate overdrive voltage of 0V (left) and \SI{-0.4}{\volt} (right). The energy is referred to the VB edge, and the Fermi level is depicted as a dashed line.}\label{fig:dos}
\end{figure}


Figure \ref{fig:bandstructure} depicts the VB of the four NWs with the different materials: from top to bottom and left to right, Si, Ge, GaSb and InSb, respectively. Two different gate overdrive voltages, $\Vg-\Vt$ (with $\Vt$ the threshold voltage), are depicted for each of these devices, 0V (left) and \SI{-0.4}{\volt} (right). Here, $\Vt$ has been evaluated as the voltage that maximizes $\dd^2 \Qi / \dd \Vg^2$ \cite{Marin2013}, where \Qi is the semiconductor charge. The depicted bandstructures energies are referred to the maximum of the VB, and the Fermi energy is represented as an horizontal dashed line. Additionally, Fig. \ref{fig:dos} plots the DOS as a function of the energy for each of the devices and biases considered in Fig. \ref{fig:bandstructure}.

The VB structure calculated with the $\kp$ method has been re-ordered using an algorithm that minimizes the changes of the subband group velocity \cite{Zlatan2012}. As a result, although the VB structure in the NW is the result of a mix between the bulk hole subbands (HH, LH and SO), Fig. \ref{fig:bandstructure} shows that we can still sort the resulting them into HH-dominated (HHD) subbands and LH-dominated (LHD) subbands, exhibiting two trends corresponding to different curvatures. HHD subbands have lower curvature and group velocity, and higher DOS than the LHD ones.


At a first glance, the Si device presents the highest DOS due to: i) the larger number of subbands per unit energy, and ii) their lower curvature, both for HHD and LHD subbands. The Ge, InSb and GaSb NWs exhibit similar bandstructures, where the HH band is much less influential, resulting in just one HHD subband and a lower DOS. When increasing the gate overdrive voltage from \SI{0}{\volt} to \SI{-0.4}{\volt}, a reduction in the energy distance between the HHD and the LHD subbands, and also between the first and second LHD subbands is observed for all materials (Fig. \ref{fig:bandstructure}) resulting in an increased DOS (Fig. \ref{fig:dos}).

\subsection{Hole distribution and centroid}

\begin{figure}
\centering
\includegraphics[width=80mm]{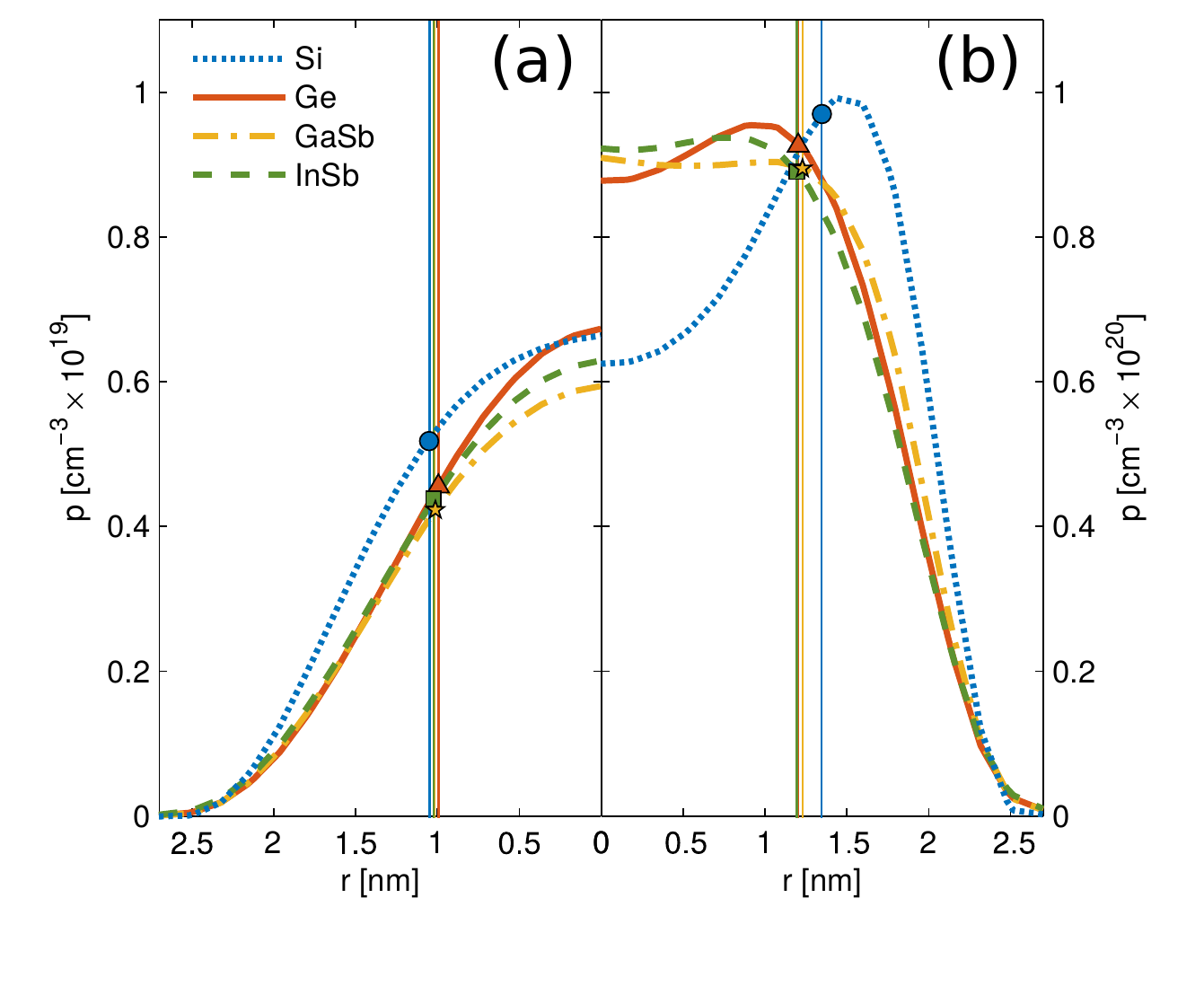}
\caption{Hole density along the radial axis for 5nm cylindrical NWs made of different materials (Si dotted blue, Ge solid red, GaSb dash-dotted orange and InSb dashed green) at a gate overdrive voltage of 0V (a) and \SI{-0.4}{\volt} (b). Vertical lines and symbols denote the position of the centroid $\centroid$ for each material: circle for Si, triangle for Ge, star for GaSb and square for InSb. $r=0$ corresponds to the center of the NW.}\label{fig:HoleDensity}
\end{figure}

Employing the bandstructure and wave functions obtained from the $\kp$ simulations, the charge distribution can be evaluated. Our results show an almost isotropic behavior of the charge for the four devices under study, due to the [111] orientation considered. Fig. \ref{fig:HoleDensity} depicts the radial hole distribution $p(r)$ in the NWs made of different materials (dotted line for Si, solid line for Ge, dash-dotted for GaSb and dashed for InSb) when two different gate overdrive voltages are considered: 0V (a) and \SI{-0.4}{\volt} (b). 

In the subthreshold and near-threshold regimes, Fig. \ref{fig:HoleDensity}(a), the charge for all the devices presents a quite similar trend, with the maximum located at the device center. Nevertheless, when the applied gate voltage increases, Fig. \ref{fig:HoleDensity}(b), the charge distribution is shifted towards the semiconductor-insulator interface, presenting a clear peak far from the device center for Si and Ge, whereas for GaSb and InSb it exhibits a broader flat hole distribution at the center of the NW. 


The displacement of the charge towards the semiconductor-insulator interface can be quantitatively evaluated using the charge centroid \cite{Lopez-Villanueva2000,Lee2012}. We will consider the logarithmic weighting of the hole distribution proposed in \cite{Lee2012}, which is closely related to the gate capacitance analyzed in the next Section. So that, the centroid ($\centroid$) of the hole distribution, referred to the semiconductor-insulator interface, is calculated as:

\begin{equation}
\centroid = R\left ( 1-\exp \left ( -2\pi \zi \right ) \right )
\end{equation}
with $\zi$ defined as:
\begin{equation}
\label{eq:zi}
\zi = \frac{1}{2\pi} \frac{\int_0^R r\,\log (\frac{R}{r})\,p(r) \dd r}{\int_0^R r\,p(r) \dd r} \eqdot
\end{equation}

Figure \ref{fig:HoleDensity} indicates the position of the centroid with a vertical line and a symbol laying on their respective curves. At low charge densities ($\Vg -\Vt = \SI{0}{\volt}$), carriers are mainly located close to the center of the device in all the cases and very similar values are found for the centroid. For large overdrive voltages ($\Vg -\Vt = \SI{-0.4}{\volt}$), the charge centroid moves closer to the interface, more markedly for Si than for the other materials.

\begin{figure}
\centering
\includegraphics[width=80mm]{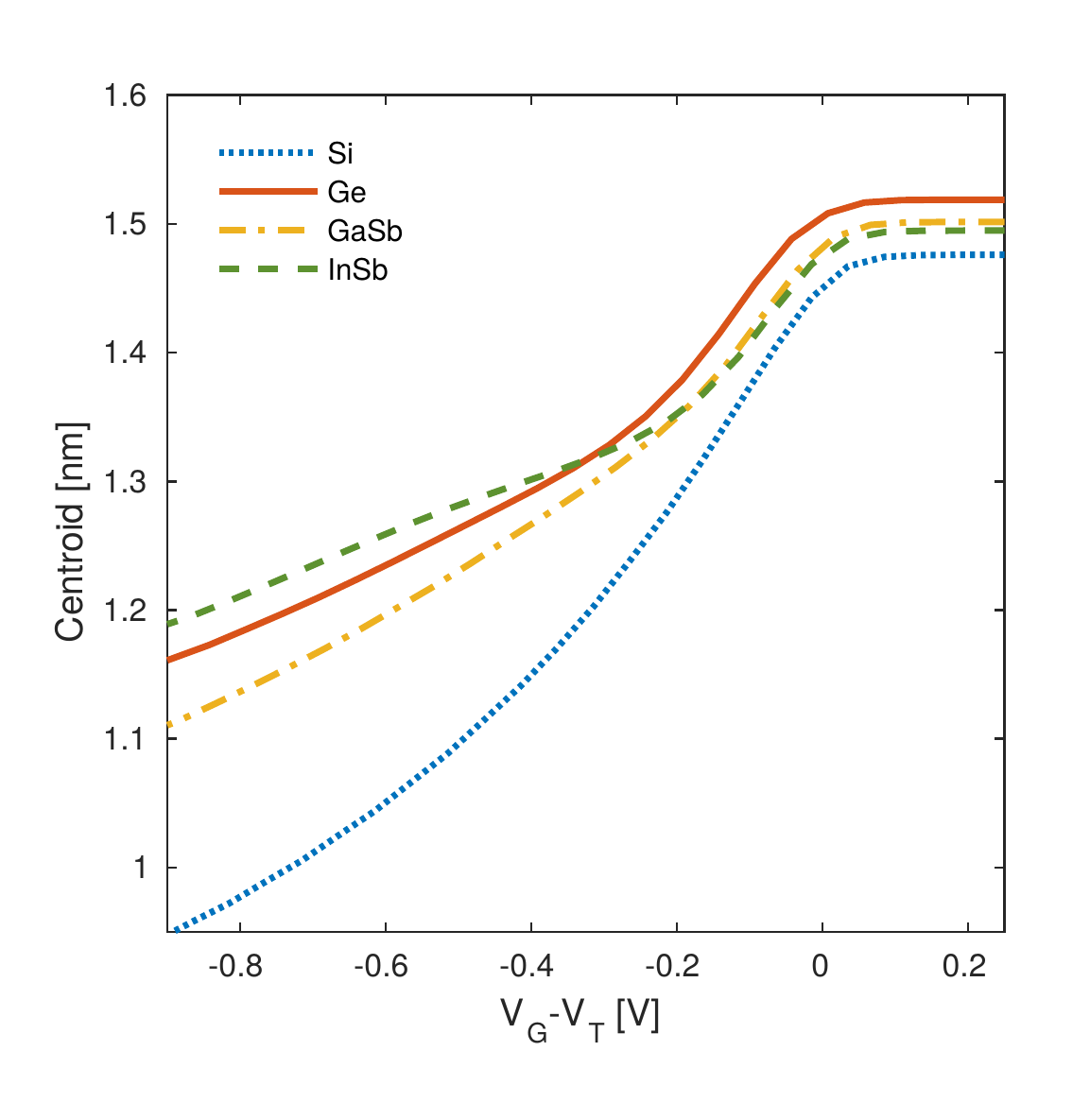}
\caption{Variation of the centroid position with respect to the semiconductor-insulator interface for the same devices as in Fig. \ref{fig:HoleDensity}.}\label{fig:centroid}
\end{figure}

Figure \ref{fig:centroid} provides deeper insight in the dependence of the centroid on $\Vg -\Vt$ for the four devices under consideration. As can be seen, the Si device has the lower value for the whole range of gate overdrive voltages analyzed. In the subthreshold regime, the centroid remains constant showing similar results regardless the material considered consistently with the charge distribution presented in Fig. \ref{fig:HoleDensity}. When the device gets into inversion, the charge centroid decreases, (as the charge is shifted towards the interface) with different trends for each material. 
Si shows the largest modulation of the centroid with $\Vg$, whilst InSb shows the smallest. With the exception of Si, a noteworthy reduction on the centroid control by the gate is observed when  $\Vg-\Vt < \SI{-0.2}{\volt}$.
As it will be shown next, this slope reduction has a direct effect on the gate capacitance of those devices.

\subsection{Gate Capacitance}
\label{sub:capacitance}

The best figure to assess the gate control over the channel charge is the gate capacitance, \Cg. With the reduction of the channel length, it has become challenging to keep high values for \Cg, being this one of the main reasons to employ GAA geometries. Thus, a comparison of this magnitude for the four devices under consideration will provide a good measure of their electrostatic performance.

For the cylindrical GAA under study, and assuming an isotropic distribution of the charge density in the cross-section, the Gauss's Law is straightforwardly applicable. Thus, $\Cg$ can be regarded as the contribution of three series capacitances: 

\begin{equation}
\label{eq:Cg}
\frac{1}{\Cg}= \frac{1}{\Cins}+\frac{1}{\Cc}+\frac{1}{\Cq}
\end{equation}

As aforementioned, the same dielectric (Al$_2$O$_3$) has been considered throughout this work, ($\Tins=\SI{1.5}{\nano\meter}$ and $\varepsilon_\text{ins}=9\varepsilon_0$). So that, $\Cins$ can be evaluated as:

\begin{equation}
\label{eq:Cins}
\Cins = \frac{2\pi \varepsilon_\text{ins}}{\ln{\left(1+\frac{\Tins}{R}\right)}}
\end{equation}

where $\varepsilon_\text{ins}$ is the insulator dielectric constant, and the rest of the parameters have been previously defined. $\Cc$ and $\Cq$ are the centroid and quantum capacitance terms, respectively, that can be evaluated as:

\begin{equation}
\label{eq:Cc}
\Cc= \frac{\partial Q \sub i}{\left( \partial \phis - \phic \right)}
\end{equation}
and
\begin{equation}
\label{eq:Cq}
\Cq= \frac{\partial Q \sub i}{\partial \phic}
\end{equation}
where $\phic$ and $\phis$ correspond to the potential evaluated at the device center and the semiconductor-insulator interface, respectively. $\Cc$ and $\Cq$ can be combined into the so-called inversion capacitance $\Cinv=\partial \Qi / \partial \phis$. $\Cq$, has been related to the finite DOS resulting from quantization \cite{Luryi1988,Marin2013}. As for \Cc, it can be calculated as \cite{Ruiz2010, Lee2012}:

\begin{equation}
\label{eq:Cc2}
\frac{1}{\Cc} = \frac{\zi}{\epsilons} +\frac{\Qi}{\epsilons} \frac{\dd \zi}{\dd \Qi}
\end{equation}

where $\epsilons$ is the semiconductor dielectric constant and $\zi$ was defined in (\ref{eq:zi}), remarking the close relation between $\Cc$ and the charge centroid \cite{Lee2012}.

\begin{figure}
\centering\includegraphics[width=85mm]{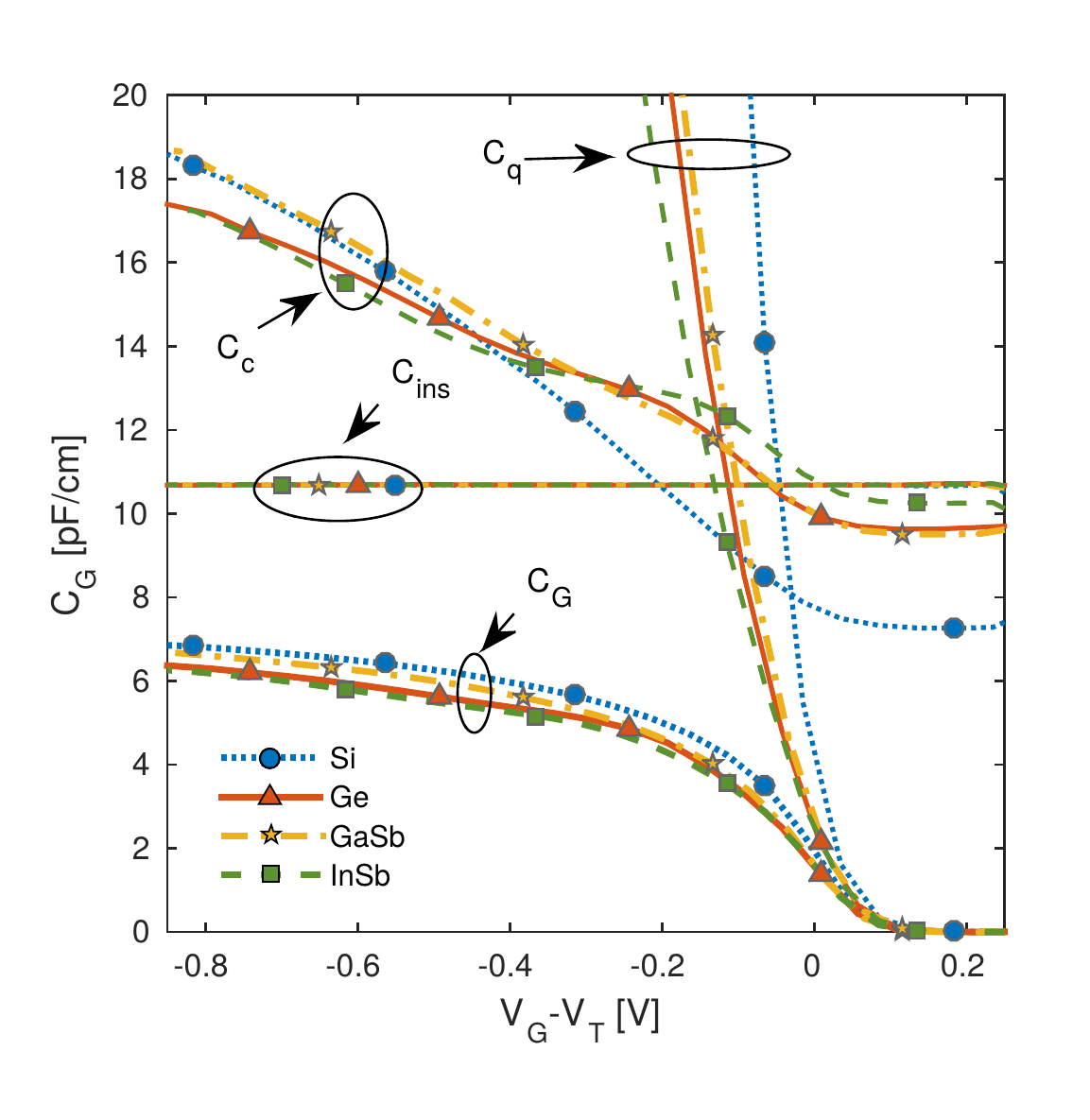}
\caption{Gate capacitance $\Cg$ for 5nm cylindrical NWs with different materials (Si in blue dotted line, Ge in red solid line, GaSb in orange dash-dotted line and InSb in green dashed line) as a function of the gate overdrive voltage $\Vg-\Vt$. The constituents capacitances $\Cins$, $\Cc$ and $\Cq$  are also depicted.}\label{fig:capacidades}
\end{figure}

In Fig. \ref{fig:capacidades}, the gate capacitances and their components are depicted for the four materials considered here. We observe a similar trend on the total gate capacitance for all of them. The Si device provides the highest value of $\Cg$ in the whole range of gate voltages but the rest of materials show a comparable performance in terms of \Cg with a maximum variation of ~15\% in strong inversion. For a better understanding of the behavior of each device, the different constituent capacitances must be analyzed. As aforementioned, $\Cins$ is the same for the four devices, so we will restrict our analysis to $\Cq$ and $\Cc$. Due to the extremely low values of $\Cq$ in the subthreshold regime, the influence of $\Cins$ and $\Cc$ in the total capacitance can be neglected in this range. For gate voltages close to $\Vt$ and slightly higher, $\Cg$ is influenced by both $\Cc$ and $\Cq$. In Si, the faster enhancement of $\Cq$, due to the larger DOS, is spoiled by a low $\Cc$. For Ge, InSb and GaSb, $\Cq$ increases more softly, remaining below \SI{20}{\pico\farad\per\centi\meter} up to $\Vg-\Vt$ around \SI{-0.2}{\volt}. This DOS bottleneck is, however, compensated by the higher $\Cc$ with respect to Si, in weak inversion. The behavior of $\Cc$ is controlled by the dielectric constant of the semiconductor and the centroid. For Si, the lower semiconductor  dielectric constant, makes the $\Cc$ term lower than for Ge, InSb and GaSb for small gate overdrives. Nevertheless, as the gate voltage becomes more negative, this effect is compensated by the smaller value of the centroid (see Fig. \ref{fig:centroid}). The centroid warping in Fig. \ref{fig:centroid} also explains the degradation of $\Cc$ in Ge, InSb and GaSb. In spite of this degradation, GaSb provides similar $\Cc$ values as compared to Si for $\Vg-\Vt \le -0.4$V. 

\begin{figure}
\centering\includegraphics[width=80mm]{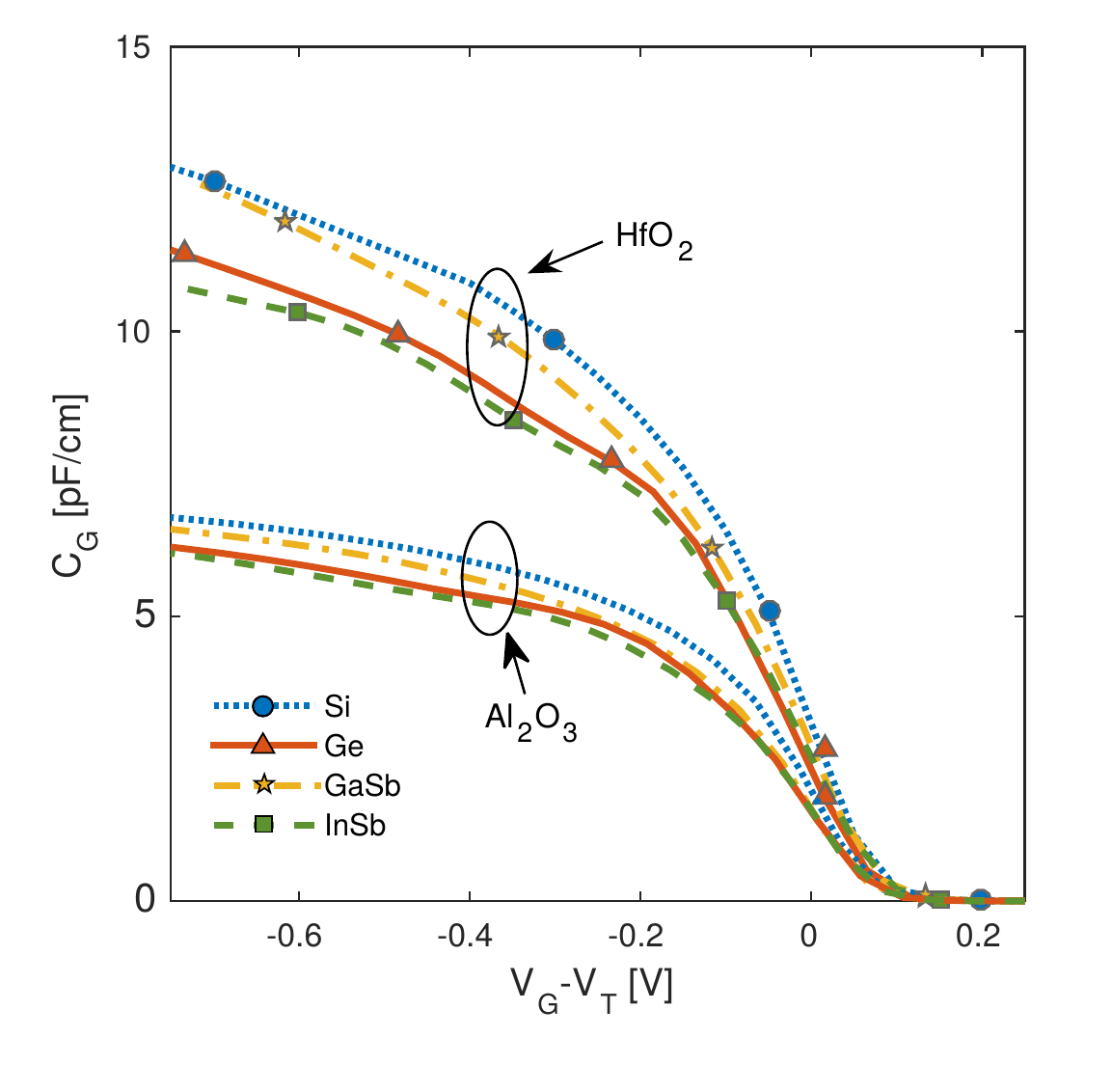}
\caption{Gate capacitance $\Cg$ for \SI{5}{\nano\meter} cylindrical NWs with two different high-$\kappa$ insulators, Al$_2$O$_3$ and HfO$_2$, with a similar thickness of \SI{1.5}{\nano\meter}. Si in blue dotted line, Ge in red solid line, GaSb in orange dash-dotted line and InSb in green dashed line as a function of the gate overdrive voltage $\Vg-\Vt$.}\label{fig:CG_hfo2}
\end{figure}

From Fig. \ref{fig:capacidades} it can be observed that the total gate capacitance is limited by the low value of $\Cins$, which veils the effect of the rest of capacitances for large overdrive voltages. This capacitance does not depend on the channel material, and can be raised by using higher $\kappa$ materials as gate insulators. New simulations have thus been carried out with HfO$_2$, a high-$\kappa$ dielectric with $\varepsilon_\text{ins}=25\varepsilon_0$ and $\Tins=\SI{1.5}{\nano\meter}$. Fig. \ref{fig:CG_hfo2} compares $\Cg$ as a function of the overdrive voltage for the four semiconductors and the two insulators, Al$_2$O$_3$ and HfO$_2$. As can be seen, $\Cg$ values achieved with HfO$_2$ double those attained with Al$_2$O$_3$. 
Again, the Si device presents the best performance in terms of gate capacitance, but GaSb holds the comparison showing an excellent electrostatics behavior in the whole range of applied bias. On the other hand, Ge and InSb NWs show similar trends with a slight reduction as $\Vg - \Vt$ becomes more negative. 
Hence, these results demonstrate the excellent electrostatic performance, in terms of gate capacitance, of the p-channel antimonide NWs considered in this study. In particular, GaSb is the most attractive alternative as it provides $\Cg$ values similar to Si in addition to the already demonstrated excellent transport properties.

\section{Conclusion} \label{sec:conclusion}
A comprehensive study of the electrostatic performance of 5nm diameter NWs has been carried out focusing on p-type channels made of four different materials: InSb, GaSb, Ge and Si. To do so, we have developed a numerical simulator that self-consistently solves the Poisson equation and the eight-band $\kp$ method in the cross section of cylindrical NWs. Our results indicate that, in spite of their lower effective mass and smaller density of states, GaSb and InSb p-type NWs hold the comparison with Si and, in the case of GaSb, outperforms Ge in terms of gate capacitance and inversion charge. Thus, their good electrostatic performance combined with the expectation of superior transport characteristics place GaSb and InSb as attractive alternatives for p-type CMOS logic based on NWs.

\bibliographystyle{IEEEtran}
\bibliography{library}

\end{document}